\documentclass[a4paper,fleqn]{cas-sc}

\usepackage[utf8]{inputenc}
\usepackage[T1]{fontenc}
\usepackage{siunitx}
\usepackage{makecell}
\usepackage{graphicx}
\usepackage{xcolor}
\usepackage{hyperref}
\usepackage{tablefootnote}
\usepackage{tikz}
\usepackage[ruled,vlined]{algorithm2e}
\usepackage{multirow}
\usepackage{threeparttable}
\usepackage{array} % Table
\usepackage{ragged2e, microtype}
\usepackage{subcaption}
\usepackage{pifont}
\usepackage{flushend}
\usepackage[numbers]{natbib}

\newcolumntype{L}[1]{>{\raggedright\let\newline\\\arraybackslash\hspace{0pt}}m{#1}}
\newcolumntype{P}[1]{>{\raggedright\arraybackslash}p{#1}}
\newcolumntype{C}[1]{>{\centering\arraybackslash}p{#1}}
\usepackage{graphics} % for pdf, bitmapped graphics files

\newcommand{\yestick}{{\color{olive}\ding{51}}}
\newcommand{\notick}{{\color{red}\ding{55}}}

\usepackage{soul}

\usepackage{etoolbox}
\newtoggle{finalPaper}

\setstcolor{red}
\togglefalse{finalPaper} % Change \toggletrue <-> \togglefalse

\iftoggle{finalPaper} {

	\newcommand{\rmvtxt}[1]{}}
{

	\newcommand{\rmvtxt}[1]{\st{#1}}}
\begin{document}

\let\WriteBookmarks\relax
\def\floatpagepagefraction{1}
\def\textpagefraction{.001}

\title[mode = title]{LwHBench: A low-level hardware component benchmark and dataset for Single Board Computers}

\shorttitle{LwHBench: A low-level hardware component benchmark and dataset for Single Board Computers}

\author[1]{Pedro Miguel {S\'anchez S\'anchez}}[orcid=0000-0002-6444-2102]
\cormark[1]

\author[1]{Jos\'e Mar\'ia {Jorquera Valero}}[orcid=0000-0003-1365-7573]

\author[2]{Alberto {Huertas Celdr\'an}}[orcid=0000-0001-7125-1710]

\author[3]{G\'er\^ome {Bovet}}[orcid=0000-0002-4534-3483]

\author[1]{Manuel {Gil P\'erez}}[orcid=0000-0002-7768-9665]

\author[1]{Gregorio {Mart\'inez P\'erez}}[orcid=0000-0001-5532-6604]

\address[1]{Department of Information and Communications Engineering, University of Murcia, Murcia 30100, Spain. Corresponding author: Pedro Miguel Sanchez (pedromiguel.sanchez@um.es)}

\address[2]{Communication Systems Group (CSG), Department of Informatics (IfI), University of Zurich UZH, 8050 Zürich, Switzerland}

\address[3]{Cyber-Defence Campus within armasuisse Science \& Technology, CH---3602 Thun, Switzerland}


\begin{abstract}
In today's computing environment, where Artificial Intelligence (AI) and data processing are moving toward the Internet of Things (IoT) and Edge computing paradigms, benchmarking resource-constrained devices is a critical task to evaluate their suitability and performance. Between the employed devices, Single-Board Computers arise as multi-purpose and affordable systems. The literature has explored Single-Board Computers performance when running high-level benchmarks specialized in particular application scenarios, such as AI or medical applications. However, lower-level benchmarking applications and datasets are needed to enable new Edge-based AI solutions for network, system and service management based on device and component performance, such as individual device identification. 
Thus, this paper presents LwHBench, a low-level hardware benchmarking application for Single-Board Computers that measures the performance of CPU, GPU, Memory and Storage taking into account the component constraints in these types of devices. LwHBench has been implemented for Raspberry Pi devices and run for 100 days on a set of 45 devices to generate an extensive dataset that allows the usage of AI techniques in scenarios where performance data can help in the device management process. Besides, to demonstrate the inter-scenario capability of the dataset, a series of AI-enabled use cases about device identification and context impact on performance are presented as exploration of the published data. Finally, the benchmark application has been adapted and applied to an agriculture-focused scenario where three RockPro64 devices are present.
\end{abstract}

\maketitle

\section{Introduction}

Performance benchmarking has been an issue explored since the early days of computer science. Knowing the capabilities of a device is critical to create applications optimized for it \cite{hockney1996science}. In this sense, benchmarking has become a priority due to the magnification of the number of online devices provoked by new technologies such as 5G, IoT or cloud. In addition, the explosion of techniques such as Machine Learning (ML) and Deep Learning (DL), which usually require high computational power, has been another key factor increasing the need for processing measurement applications. In this context, \textit{device performance benchmarking} is a research avenue that acquired a large momentum in the last years \cite{varghese2021survey}, especially in IoT and Edge computing paradigms. Benchmarking can be defined as an intentional stress introduced in a system to measure how the device behaves regarding a determined set of metrics \cite{john2018performance}. The main purposes of benchmarking are to \textit{i)} verify that the performance of a device is the promised one and suitable for a certain activity such as running AI tasks, and \textit{ii)} model the internal behavior of a certain device to identify it or verify that it is running properly \cite{sanchez2021survey}. Therefore, device benchmarking can be seen from two perspectives. First, as a high-level system benchmarking that seeks to measure how a certain application works on the device in terms of performance or Quality-of-Service (QoS), for example, in execution time or energy consumption. Second, as a low-level hardware benchmarking, which goal is to characterize the device components in a more precise way \cite{wyant2012computing}, so it is possible to differentiate devices or detect imperfections and errors in the chips, among other options. ``Low-level'' indicates that the focus of the measurements is close to the physical level of the components, measuring values such as frequencies and cycles instead of QoS (time and energy). An example of this type of benchmarking would be to measure how many cycles it takes a processor to execute a certain simple task, measuring whether it meets its specification and the necessary stability requirements.

Moreover, due to the explosion undergone by AI, and more specifically ML and DL \cite{dong2021survey}, these techniques have also landed in the IoT field, being applied in areas such as industry, network and service management or cybersecurity. Thus, the field of IoT benchmarking also has the task of evaluating the performance of training and deploying ML/DL in these devices hardware. Numerous solutions have explored the capabilities of IoT devices, mainly Single-Board Computers (SBC) such as Raspberry Pi, when running different ML/DL algorithms and libraries \cite{zhang2018pcamp}, so this is an area with enough pedigree and many recent works \cite{baller2021deepedgebench}. However, low-level hardware benchmarking and the application of ML/DL in the data generated from these benchmarks remain mostly unexplored in the IoT field. This is an important task as critical functionalities are moving to the IoT and having the components of these devices properly analyzed is essential.

Then, although many benchmarks have been proposed for SBCs in recent years, as \cite{varghese2021survey} shows, some challenges are present in the area, such as \textit{(i)} many benchmarks are proposed but no exhaustive execution datasets are provided; \textit{(ii)} all recent benchmarks focus on high-level applications and none of them measures the performance of the hardware from a low-level perspective; \textit{(iii)} there is no work measuring the components performance from a different component of the same device, which is important to avoid inconsistent values coming from the measured component (e.g. measure an execution time using the CPU as its own source of timestamps), as it can not notice its own inaccuracy; \textit{(iv)} only a few solutions consider storage and memory in the benchmark, most of them are focused on CPU QoS (execution time, computation and communication latency); and \textit{(v)} none of the previous SBC benchmark solutions consider GPU low-level performance for non-graphics processing.

In  order to improve the previous limitations and fill the literature gap, the main contributions of the present work are:
\begin{itemize}
    \item A low-level hardware component benchmarking application, namely LwHBench, which measures CPU, GPU, memory and storage device performance from the device reference point. The benchmark is implemented as a proof of concept for Raspberry Pi devices \cite{code}, taking into account the particularities of their hardware components.
    
    \item A comprehensive dataset acquired as a result of running the previous benchmark on a set of 45 Raspberry Pi of various models for 100 days \cite{dataset}. For data collection, a series of measures regarding the stability of the device performance have been taken, setting the frequency of the components to fixed values and trying to reduce as much as possible the possible noise introduced in the measurements by other processes running on the devices. This dataset contains a total of 4 GB of data (2386126 vectors), more than any other benchmark dataset in the literature, ready to be used in ML/DL-based applications by other researchers in a wide variety of use cases. 
    
    \item A set of potential use cases described as possible application scenarios for the benchmark and the published dataset. These use cases are partially solved as a preliminary exploration of the dataset, so that other researchers know how to apply their ML/DL algorithms. The code used in these use cases is also publicly available at \cite{code}. Besides, these use cases are also demonstrated in a real world IoT agriculture deployment using 3 RockPro64 devices, another SBC model. 
    
%    \item An additional implementation of the benchmarking application for Mali GPU-based devices, which serves as a guideline for other researchers who wish to adapt the benchmarking application to new SBC models.

\end{itemize}

The remainder of this article is structured as follows. Section \ref{sec:related} provides a review of the literature status regarding SBC performance benchmarking. Section \ref{sec:methodology} describes the methodology and approach followed for the benchmark implementation and generation of the dataset. Section \ref{sec:benchmark} describes the functions executed in each one of the components considered, while Section \ref{sec:exploration} explores the collected data through a set of use cases. Section \ref{sec:deployment} depicts a real-world adaptation and deployment of the benchmark. Section \ref{sec:dicussion} draws the main strong points of the proposed method together with the drawbacks identified during the work development. Finally, Section \ref{sec:conclusion} draws the main conclusions of the present work and future research lines.

\section{Related Work}
\label{sec:related}

This section analyzes related work dealing with performance benchmarking, with a special focus on Edge computing and IoT, and existing datasets regarding IoT device performance monitoring.

\begin{table*}[]
\centering
\caption{Comparison of IoT Benchmarking applications and datasets.}
\label{tab:comparison}
%\resizebox{\textwidth}{!}{%
\scriptsize
\begin{tabular}{cllllll}
\makecell[c]{\textbf{Solution}} & \makecell[c]{\textbf{Device Type}} & \makecell[c]{\textbf{Monitored Components}} & \makecell[c]{\textbf{Metrics}} & \makecell[c]{\textbf{Public}\\\textbf{Data}} & \makecell[c]{\textbf{Open}\\\textbf{Source}} \\ \hline   \hline

\cite{grasso2014energy} & SBC & GPU & Processing time, resource usage (\%) & \notick & \notick \\
\hline
\cite{cloutier2016raspberry} & SBC & CPU & GFLOPS, energy & \notick & \notick \\
\hline
\cite{zhang2018pcamp} & SBC / PC & CPU, GPU, memory & Processing time, mem speed, energy & \notick & \yestick \\
\hline
\cite{limaye2018hermit} & Medical IoT & CPU, memory & Hardware Performance Counters & \notick & \yestick \\
\hline
\cite{das2018edgebench} & Cloud & CPU, memory, network & Processing time, network / mem speed & \notick & \notick \\
\hline
\cite{mcchesney2019defog} & SBC & CPU, memory, network & Processing time, network / mem speed & \notick & \yestick \\
\hline
\cite{klervie_tocze_2020_3974220} & Edge Servers & CPU & N Instructions & \yestick (2.5 MB) & \yestick \\
\hline
\cite{hawthorne2020evaluating} & SBC & CPU & GFLOPS & \notick & \yestick \\
\hline
\cite{gatech-fingerprinting-20140609} & IoT & CPU & Packet inter-arrival time & \yestick (1.5 GB) & \notick \\
\hline
\cite{hagelskjaer2020dataset} & IoT & Radio transmitters & Raw transmission data & \yestick (+50 GB) & \notick \\
\hline
\cite{basford2020performance} & SBC & CPU & GFLOPS, energy & \notick & \yestick \\
\hline
\cite{yang2021noisfre} & Any & SRAM and Flash memories & Initial bit status & \yestick (To be published) & \yestick \\
\hline
\textbf{\makecell[c]{This\\work}} & \textbf{SBC} & \textbf{CPU, GPU, Memory, Storage} & \textbf{Cycles counters and processing time} & \yestick \textbf{(4 GB)} & \yestick \\
\hline
\end{tabular}%
%}
\end{table*}

Regarding performance benchmarking, Varghese et al. \cite{varghese2021survey} surveyed the evolution of this field from the early 90s to 2020, with special consideration of Edge benchmarking since the 2010s. This survey shows that many performance benchmarking applications have been published in recent years, most of them centered on SBCs, such as Raspberry Pi. The vast majority of these applications focus on CPU and memory benchmarking \cite{limaye2018hermit}, while only a few test additional resources \cite{ramalho2016virtualization}, such as storage, network or accelerators (GPU/TPU). Despite the usefulness to test devices from different brands and models, none of the existing Edge performance benchmarks revised in \cite{varghese2021survey} is focused on the extraction of low-level information capable of detecting hardware imperfections or malfunctioning. In contrast, these existing benchmarks are based on the execution of complex or advanced applications \cite{mcchesney2019defog}, such as AI libraries \cite{zhang2018pcamp} or orchestrators, and not on fast execution code for low-level fingerprinting. Some benchmarks \cite{das2018edgebench} also measure the performance of cloud platforms oriented to the IoT, such as AWS Greengrass or Azure IoT Edge. Regarding measurement metrics, most of the benchmarks use time-based metrics such as GFLOPS (Giga Floating Point Operations per Second) for CPU or MB/s for memory and network, with only a few of them using more complex and low-level ones such as Hardware Performance Counters \cite{limaye2018hermit}. Besides, although most benchmarking applications provide datasets with them (12 of 14 analyzed applications), only 5 out of 14 benchmarking applications use full open-source software, while 10 out of 15 use commercial-grade software. Finally, \cite{varghese2021survey} also shows that most high-level benchmarks use commercial or proprietary software in their implementations.

In the area of low-level benchmarking, \cite{grasso2014energy} evaluated the performance of the GPUs embedded in ARM SBCs, noticing great improvements comparing the GPU performance to the CPU when doing mathematical operations, but with higher energy consumption. Furthermore, \cite{cloutier2016raspberry} built a Raspberry Pi cluster and performed CPU and energy consumption testing to find the best energy/price/performance model. Similarly, three clusters, each consisting of 16 different SBC models, were built in \cite{basford2020performance} to benchmark the SBC performance in terms of computing and energy consumption. \cite{hawthorne2020evaluating} followed a similar cluster-oriented benchmarking but focusing on cryptography libraries. However, these benchmarks perform fairly simple metrics about performance and do not publish their data, which do not enable the application of the generated data to new domains such as fingerprinting. From a different domain, \cite{wright2019benchmarking} explored recently the low-level benchmarking of quantum computers, an area gaining importance in recent years that supports the need for lower-level hardware benchmarking.

Dealing with datasets about hardware performance, there are just a few examples available in the literature. Many benchmarking applications include simple data samples \cite{varghese2021survey}. One example is \cite{klervie_tocze_2020_3974220}, which contains 2.5 MB of traces of different high-level benchmarking applications executed in Edge servers. However, these are not exhaustive datasets collected during long execution periods and are not suitable for ML/DL approaches due to their size constraints. Regarding datasets directly focused on low-level performance fingerprinting, \cite{yang2021noisfre} contains fingerprints from different SRAM (Static RAM) chips, which were used in \cite{gao2021noisfre} to perform individual identification. However, most SBC models do not include SRAM chips due to their higher cost. From a different perspective, \cite{hagelskjaer2020dataset} contains radio spectrum measurements from different IoT devices, which can be employed to fingerprint their transmission performance and properties. Similarly, \cite{gianmarco2020iot} also contains raw IQ signals from 9 IoT devices that can be used for fingerprinting tasks. Moreover, \cite{gatech-fingerprinting-20140609} presents inter-arrival time information from different wireless routers and IoT devices, and it is aimed at individual and device type fingerprinting. In contrast, to the best of our knowledge, there is not any comprehensive dataset regarding low-level performance fingerprinting or benchmarking of hardware components.

\tablename~\ref{tab:comparison} shows a comparison between the different benchmarking applications and datasets found in the literature and the present one. From the analysis made in this section, it is noticed that there is a gap regarding solutions focused on low-level benchmarking. Most of the recent solutions focus on high-level application benchmarking, and the ones focusing on low-level performance only use simple performance metrics and do not provide extensive datasets to enable ML/DL-based use cases or new applications. Moreover, the datasets found in the literature are focused on other areas such as device identification, and not on hardware component benchmarking.

\section{Benchmark and Dataset Generation Methodology}
\label{sec:methodology}

This section describes the methodology followed in order to implement the benchmark application and collect the samples available in the dataset: providing the details of the scenario used for data collection; describing the components monitored and how their performance is measured; detailing the libraries used to collect each metric; and finally, explaining the configuration options and measures taken to ensure stability and avoid noise in the samples published.

\subsection{Deployment and Configuration}

For the LwHBench benchmark implementation and testing, a wide number of devices is required. In this sense, the benchmark is executed, to collect the dataset, in a set of 45 physical devices composed of several models of Raspberry Pi (RPi) devices. \tablename~\ref{tab:devices} shows a summary of the devices employed for validation, their distribution and main characteristics.

\begin{table*}[htpb]
    \centering
    \scriptsize
    \caption{Number, model and most relevant characteristics of the devices used for validation.}
    \resizebox{\textwidth}{!}{
    \begin{tabular}{ >{\Centering}m{0.7cm}  >{\Centering}m{1.7cm} >{\Centering}m{1.1cm} >{\Centering}m{1.1cm} >{\Centering}m{2.5cm} >{\Centering}m{1.5cm} >{\Centering}m{3.5cm} >{\Centering}m{1cm} >{\Centering}m{1.2cm}}
        
         \textbf{N} & \textbf{RPi model}& \textbf{Revision} & \textbf{SoC} & \textbf{CPU} & \textbf{GPU} & \textbf{Cache} & \textbf{RAM} & \textbf{SD}\\
        \hline \hline
        15 & Raspberry Pi 4 Model B & 1.1/1.4 & BCM2711 & 1.5 GHz quad-core 64 bit ARM A72 & 500 MHz Broadcom VideoCore VI & 2-way set associative 32 kB and 48 kB level one instruction and data caches, respectively, and a 1 MB unified level two cache & 4 GB LPDDR4& 16GB A1 Type 10 \textit{SandDisk Ultra}\\
        \hline
        10 & Raspberry Pi 3 Model B+ & 1.3 & BCM2837 & 1.4 GHz quad-core 64 bit ARM A53 & 400 MHz Broadcom VideoCore IV & 2-way set associative 16 kB level one instruction and data caches, and 512 kB unified level two cache & 1 GB LPDDR2 & 16GB A1 Type 10 \textit{SandDisk Ultra}\\
        \hline
        10 & Raspberry Pi Model B+ & 1.2 & BCM2835 & 700 MHz single-core 32 bit ARM 1176JZF-S & 400 MHz Broadcom VideoCore IV & 2-way set associative 16 kB level one instruction and data caches, and 128 kB unified level two cache & 500 MB LPDDR2& 16GB A1 Type 10 \textit{SandDisk Ultra}\\
        \hline
        10 & Raspberry Pi Zero & 1.3 & BCM2835 & 1 GHz single-core 32 bit ARM 1176JZF-S & 400 MHz Broadcom VideoCore IV & 2-way set associative 16 kB level one instruction and data caches, and 128 kB unified level two cache & 500 MB LPDDR2& 16GB A1 Type 10 \textit{SandDisk Ultra}\\
        \hline
    \end{tabular}}
    \label{tab:devices}
\end{table*}

%\begin{itemize}
 %   \item \textit{15 Raspberry Pi 4 Model B Revision 1.1/1.4}. The System-on-Chip (SoC) on these devices is a Broadcom BCM2711, including a 1.5 GHz quad-core 64 bit ARM A72 CPU and a 500 MHz Broadcom VideoCore VI GPU. They also include 2-way set associative 32 kB and 48 kB level one instruction and data caches, respectively, and a 1 MB unified level two cache. The devices for the data collection also included 4 GB LPDDR4 RAM.
    
  %  \item \textit{10 Raspberry Pi 3 Model B+ Revision 1.3}. These devices include a  BCM2837 SoC featuring a 1.4 GHz quad-core 64 bit ARM Cortex A53 CPU and a 400 MHz Broadcom VideoCore IV GPU. The caches are 2-way set associative 16 kB level one instruction and data caches, and 512 kB unified level two cache. Regarding memory, they have 1 GB LPDDR2 RAM.
    
   % \item \textit{10 Raspberry Pi Model B+ Revision 1.2}. These devices include a  BCM2835 SoC featuring a 700 MHz single-core 32 bit ARM 1176JZF-S CPU and a 400 MHz Broadcom VideoCore IV GPU. The caches are 2-way set associative 16 kB level one instruction and data caches, and 128 kB unified level two cache. Regarding memory, they have 500 MB LPDDR2 RAM. 
    
%    \item \textit{10 Raspberry Pi Zero Revision 1.3}. These devices include a BCM2835 SoC featuring a 1 GHz single-core 32 bit ARM 1176JZF-S CPU and a 400 MHz Broadcom VideoCore IV GPU. The caches are 2-way set associative 16 kB level one instruction and data caches, and 128 kB unified level two cache. Regarding memory, they have 500 MB LPDDR2 RAM. 
%\end{itemize}

All devices of the setup have identical software images, using \textit{Raspbian 10 (buster) 32 bits} as OS and \textit{Linux kernel 5.4.83}. The only variation in the kernel version is related to the core architecture of each device model, \textit{ARMv6} for RPi1/Zero, \textit{ARMv7} for RPi3, and \textit{ARMv8} for RPi4. 

Besides, to reduce physical context as much as possible, all devices are located in the same lab room with identical cases and aluminum heat sinks.

\subsection{Monitored Components}

For reliable time/performance measurements of hardware components, the ideal setup is to use as reference a physical oscillator independent of the component being measured (the frequency of the component is not dependent on the reference oscillator). RPi devices include all the main processing components present in a normal computer, including oscillators. However, the number of physical crystal oscillators is reduced to save costs. RPi4 only includes a SoC base oscillator running at 54 MHz and an oscillator for the USB/Ethernet controllers running at 25 MHz. In contrast, RPi3/1/Zero only include one SoC base oscillator running at 19.2 MHz. Then, each component runs at a different frequency using Phase-Locked Loops (PLLs) for base frequency multiplication \cite{undocumentedPi}. 

This condition implies that some auxiliary components which could be used as reference points, such as the Real-Time Clock (RTC), are simulated. This fact makes it hard to accurately measure the performance and skew of the system from the device itself, as a skew in the CPU timing will affect the time measurements that it is doing of itself.

However, each component can still show performance differences based on the multiplication factor applied to the base crystal oscillator frequency in the associated PLLs. For this reason, each component used to measure the performance of the device is monitored from another component of the device, measuring, in turn, the possible imperfections and deviations between the components. More clearly, for example, the performance of code execution on the CPU is measured in terms of GPU cycles and vice versa.

Following the previous approach, the components whose performance are monitored and stored in the dataset are:

\begin{itemize}
    \item \textit{CPU}. The execution time of code run in the CPU is measured by monitoring how many GPU cycles have elapsed during that period of time. In this way, the skew between CPU and GPU can be accurately measured. Therefore, the formula to measure CPU performance based on GPU cycle variation is:
    \begin{equation}
        CPU_{perf.} = \Delta GPU_{cycle\_counter} \vdash t_{cpu\_code\_exec}
    \end{equation}
    
    \item \textit{GPU}. In the GPU, the performance of code executed in this component is measured using CPU-based time, just the opposite way to the previous case.
        \begin{equation}
        GPU_{perf.} = \Delta CPU_{cycle\_counter} \vdash t_{gpu\_code\_exec}
    \end{equation}
    \item \textit{Memory}. Here, memory read/write operations are also monitored in terms of CPU-based timing, as the RAM chip in RPi has its own functioning frequency different to both the CPU and GPU.
    \begin{equation}
        Memory_{perf.} = \Delta CPU_{cycle\_counter} \vdash t_{mem\_ops}
    \end{equation}
    \item \textit{Storage}. Storage performance measurement is done by input/output operations in the SD card attached to the device with the software image.
    \begin{equation}
        Storage_{perf.} = \Delta CPU_{cycle\_counter} \vdash t_{io\_ops}
    \end{equation}
\end{itemize}

\subsection{Benchmark Implementation}

The LwHBench benchmark code is available in \cite{code}. The data collection program has been implemented using Python 3, so it can be executed as a portable script on any device with the required libraries installed. Besides, this selection has also been influenced by the set of libraries available for CPU and GPU low-level interaction, as it is explained below.

For the GPU low-level interaction and register monitoring, Idein py-videocore \cite{py-videocore} and py-videocore6 \cite{py-videocore6} are employed for VideoCore IV and VideoCore VI GPU-based devices, respectively. To measure GPU cycles, different registers should be considered depending on the GPU version. In particular, the register monitored in VideoCore VI is \textit{CORE\_PCTR\_CYCLE\_COUNT}, while in VideoCore IV the monitored registers are the \textit{performance counters 13-19} \cite{videocoreIV}.

When accessing the CPU cycle counter, there are two different possibilities. The first is to generate a kernel module to enable reading the cycle counter from \textit{userspace}. The second consists of using the interfaces provided by performance monitoring tools such as \textit{perf}. In the code, both approaches are tested. 

Regarding the kernel module, the CPU cycle counter is stored in ARMv7 and ARMv8 AArch32 processors using \textit{c15 Cycle Counter Register (CCNT)} and, by default, it can only be read in kernel space. Therefore, a custom kernel module is implemented to allow access to this register (\textit{enable\_ccr} folder in \cite{code}). Once compiled, the kernel module should be loaded using \textit{insmod} command with \textit{root} privileges. Finally, the register can be read using the assembly operation \textit{MRC p15, 0, <Rd>, c15, c12, 1}, where \textit{<Rd>} represents the variable where to store the register value. To access \textit{perf} time counting, the easiest method is to use the \textit{time} built-in Python library, which includes \textit{perf\_counter\_ns()} function to retrieve time in nanoseconds using the previous counter. After some experimentation, it was decided to follow the \textit{perf}-based approach due to its simplicity compared to using a kernel module and assembly code, and its similar consistency in the performance measurements.

Moreover, to automatize the data collection process, a system service has been implemented (\textit{data\_collection.service} in the code folder). This service is in charge of the automated data collection script launching and periodic system rebooting in order to reduce the noise introduced by possible factors related to the system running time. Concretely, each device is rebooted after 800 samples are collected.

\subsection{Device Setup for Component Stability and Isolation}

One of the most critical aspects of collecting reliable samples is to ensure that the conditions in the device are as constant as possible, reducing potential sources of noise in the samples. To this end, a number of measures are taken to counteract the impact of other processes running on the device. Specifically, the measures implemented to ensure stability are:

\begin{itemize}
    \item \textbf{Fixed CPU/GPU/RAM frequency}. By default, the kernel dynamically manages the frequency of the device components to save energy when no high task load is present. However, this dynamicity affects to the stability of the performance measurements. Therefore, a fixed frequency is required in the components to measure. In RPi, the frequency of the components can be set to be constant at the maximum using \textit{turbo\_mode=1} boot option. Besides, if only a fixed CPU frequency is wanted, \textit{performance} can be used as \textit{scaling\_governor} option. 
    
    \item \textbf{Kernel level priority}. Enabling a high priority for the data collection process minimizes the interruptions caused by other programs, removing noise and inaccurate measurements. The best option here is to set the process with the highest scheduling priority. If root privilege is available, using the command \texttt{chrt --rr XX} when launching the program enables the ``real-time scheduling'' of the process, just like a kernel process. If it is not possible to use kernel priority, another option is to use \textit{nice -n -20} to set the maximum \textit{user level} priority.
    
    \item \textbf{Disable Memory Address Space Layout Randomization (ASLR)}. Memory random address organization can affect the stability of memory-related measurements; therefore, this characteristic was disabled during data collection. It can be done using \texttt{sysctl kernel.randomize\_va\_space=0} command, but note that this should be only enabled during memory-related data acquisition, as having ASLR disabled increases the facility to perform memory-based attacks such as buffer overflows.
    
    \item \textbf{Profiled Guided Optimization (PGO)}. PGO is a compiler option intended to improve runtime performance based on static program analysis of code. Python interpreter can be compiled to use PGO by using \textit{--enable-optimizations} option. Furthermore, \textit{Python garbage collector} is also disabled to avoid unintended tasks during the execution.
    
    \item \textbf{Fixed hash seed}. As hash-based CPU performance measurements are generated, using a fixed seed improves the deterministic characteristics of the function. This is set using \textit{PYTHONHASHSEED=0} (or any other number) as environment variable when running the data collection script. This option should only be used for benchmark, as it can lead to an attacker causing a Denial-of-Service (DoS) by using worst-case performance inputs to the function, which have $O(n^2)$ complexity.
    
    \item \textbf{Core isolation}. For the multi-core CPU devices, e.g. RPi3 and RPi4, one core is isolated from the rest to execute the benchmark on it using \textit{cpu affinity}. This setup avoids the (kernel) interruptions caused by other processes running in the same CPU core while the data is being generated. Concretely, the kernel options employed were: \textit{i) isolcpus}, to avoid the kernel to schedule any process in that core; \textit{ii) nohz\_full}, to tell the kernel to remove as much kernel noise as possible, such as tick interrupts; and \textit{iii) rcu\_nocbs}, to offload Read-Copy-Update (RCU) threads and callbacks.
\end{itemize}

\section{LwHBench Benchmark and Dataset}
\label{sec:benchmark}

This section details the operations executed by the benchmark and the events collected as features for the generation of the dataset associated with this paper. Note that this list can be updated by any other researcher just changing the code of the benchmarking script \cite{code}.

As detailed in Section \ref{sec:methodology}, the components leveraged for benchmarking are: CPU, GPU, Memory and Storage. They have been selected because they are the most common hardware elements in any SBC (and generic computer). \tablename~\ref{tab:features} shows the list of features collected for each device component. As it can be appreciated, any of the operations measures typical processing power metrics such as GFLOPS. This is because they are not well suited for low-level component characterization. Besides, they have already been gathered in several previous studies, as shown in Section \ref{sec:related}. Concretely, the operations implemented as proof of concept are:
\begin{itemize}
    \item \textit{CPU}. Different sleep times (from 1 to 120 seconds) are monitored, trying to measure the accuracy for time keeping in the component. Additionally, some quick-execution functions are monitored: hash calculation of a string, pseudorandom number generation, random number generation using \textit{/dev/urandom} interface, and Fibonacci number calculation.
    
    \item \textit{GPU}. Three simple operations are monitored in terms of CPU time: a matrix multiplication, a matrix summation and the processing of a graphic shadow.
    
    \item \textit{Memory}. The operations executed are the generation of a list object with 1000 integers, the reserve of 100 MB of data and the time to read a 500 kB \textit{csv} file. These operations are measured also in terms of CPU time and represent 3 features in the generated data vector.
    
    \item \textit{Storage}. 100 read and write operations of 100 kB of data are monitored in the device SD card, generating 200 features in total.
\end{itemize}

In the dataset \cite{dataset}, each value of \tablename~\ref{tab:features} represents one feature in the vectors (each dataset entry). Besides, each vector ends with the MAC address of the device, which can be used as the label in supervised ML/DL tasks. The data of each device is stored in a \textit{csv} file, whose name is the MAC address. Additionally, a text file named \textit{MAC-Model.txt} contains the association between the MAC and the model of each device in the testbed.

\begin{table}[htpb!]
    \centering
    \caption{Features gathered during data collection.}
    \begin{tabular}{ P{1.9cm} P{2.1cm} P{11cm} } 
    \textbf{Component} & \textbf{Function} & \textbf{Monitored Feature} \\
    \hline
    \hline
    \textbf{-} & timestamp & Unix timestamp \\
    & temperature & Device core temperature \\
    \hline
    \textbf{CPU} & 1 s sleep & GPU cycles elapsed during 1 second CPU sleep \\
     & 2 s sleep & GPU cycles elapsed during 2 seconds CPU sleep \\
     & 5 s sleep & GPU cycles elapsed during 5 seconds CPU sleep \\
     & 10 s sleep & GPU cycles elapsed during 10 seconds CPU sleep \\
     & 120 s sleep & GPU cycles elapsed during 120 seconds CPU sleep \\
     & string hash & GPU cycles elapsed during a fixed string hash calculation \\
     & pseudo random & GPU cycles elapsed while generating a software pseudo-random number \\
     & urandom & GPU cycles elapsed while generating 100 MB using \textit{/dev/urandom} interface \\
     & fib & GPU cycles elapsed while calculating Fibonacci number for 20 using the CPU \\
     \hline
    \textbf{GPU} & matrix mul & CPU time taken to execute a GPU-based matrix multiplication \\
     & matrix sum & CPU time taken to execute a GPU-based matrix summation \\
     & scopy & CPU time taken to execute a GPU-based graph shadow processing \\
     %&  &  \\
    \hline
     \textbf{Memory} & list creation & CPU time taken to generate a list with 1000 elements \\
     & mem reserve & CPU time taken to fill 100 MB in memory \\
     & csv read & CPU time taken to read a 500 kB \textit{csv} file \\
    \hline
    \textbf{Storage} & read x100 & 100 CPU time measurements for 100 kB storage read operations \\
     & write x100 & 100 CPU time measurements for 100 kB storage write operations \\
    \hline
    \end{tabular}
    \label{tab:features}
\end{table}

\begin{figure}[htpb!]
    \centering
    \includegraphics[width=0.45\columnwidth]{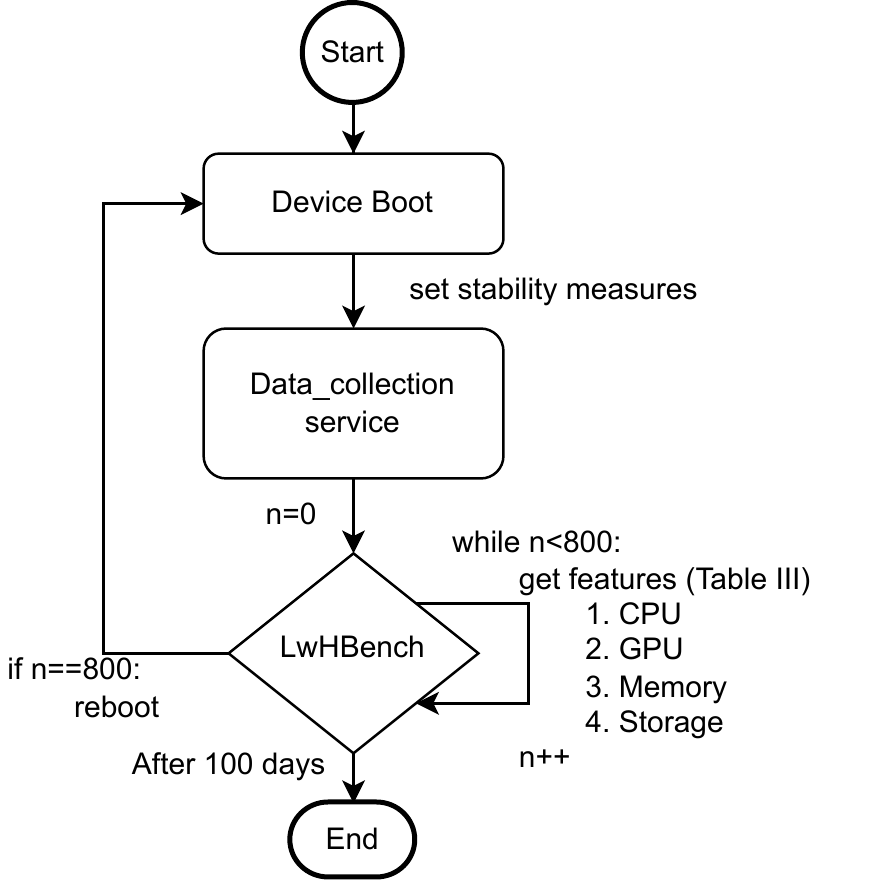}
    \caption{LwHBench data collection flow diagram.}
    \label{fig:flow}
\end{figure}

For data collection purposes, an additional Linux service has been developed (\textit{data\_collection.service)}. It is in charge of launching LwHBench when the device is booted and it takes care of rebooting the device once 800 samples have been collected. This reboot is done to minimize the possible impact of running time in the collected data (e.g. memory usage of persistent processes). \figurename~\ref{fig:flow} shows the flow diagram of the data collection performed in each device.

\begin{figure}[htpb!]
    \centering
    \includegraphics[width=0.5\columnwidth]{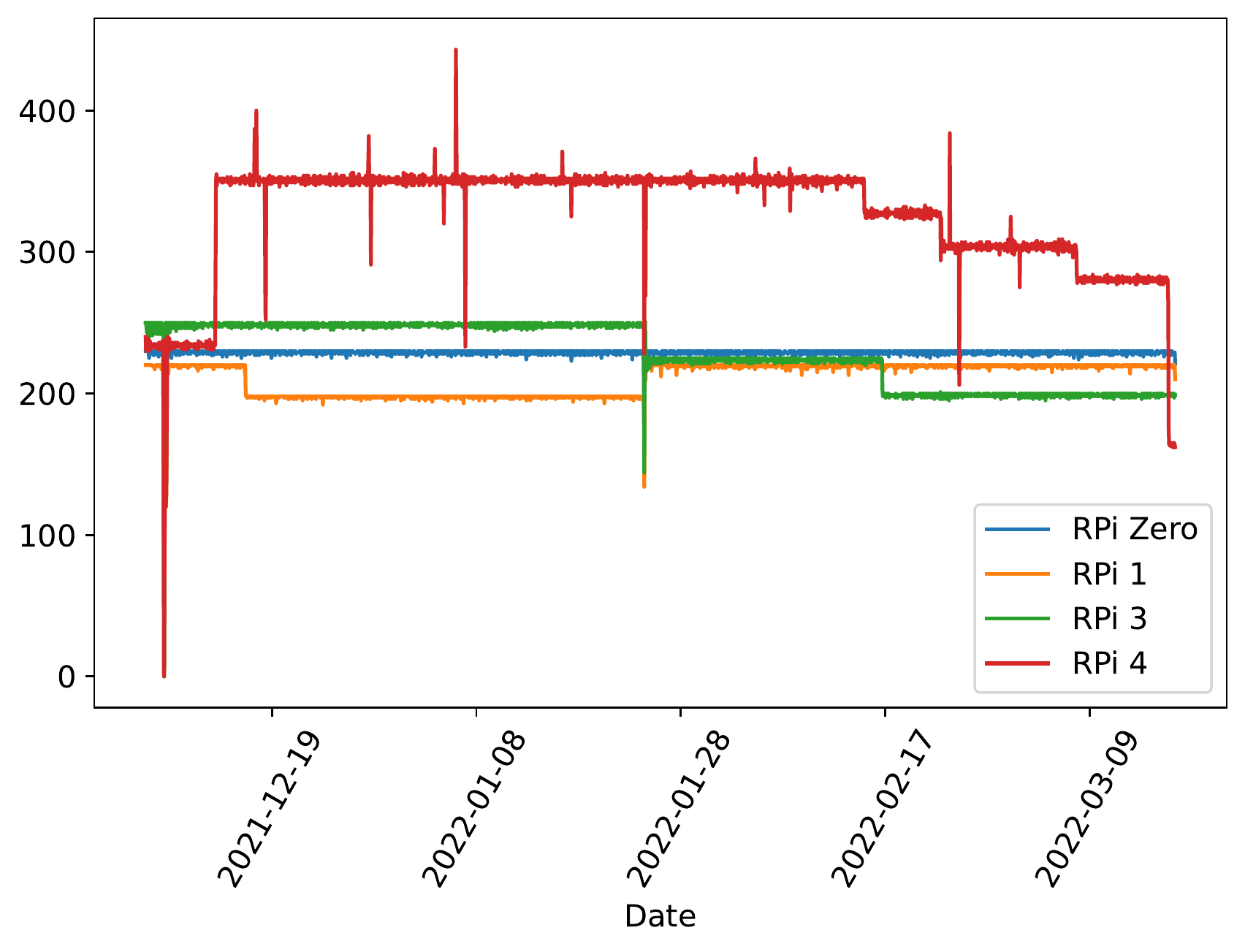}
    \caption{Samples per hour generated from each device model.}
    \label{fig:hour_hist}
\end{figure}

\begin{figure*}[htpb!]
    \centering
    \includegraphics[width=\textwidth]{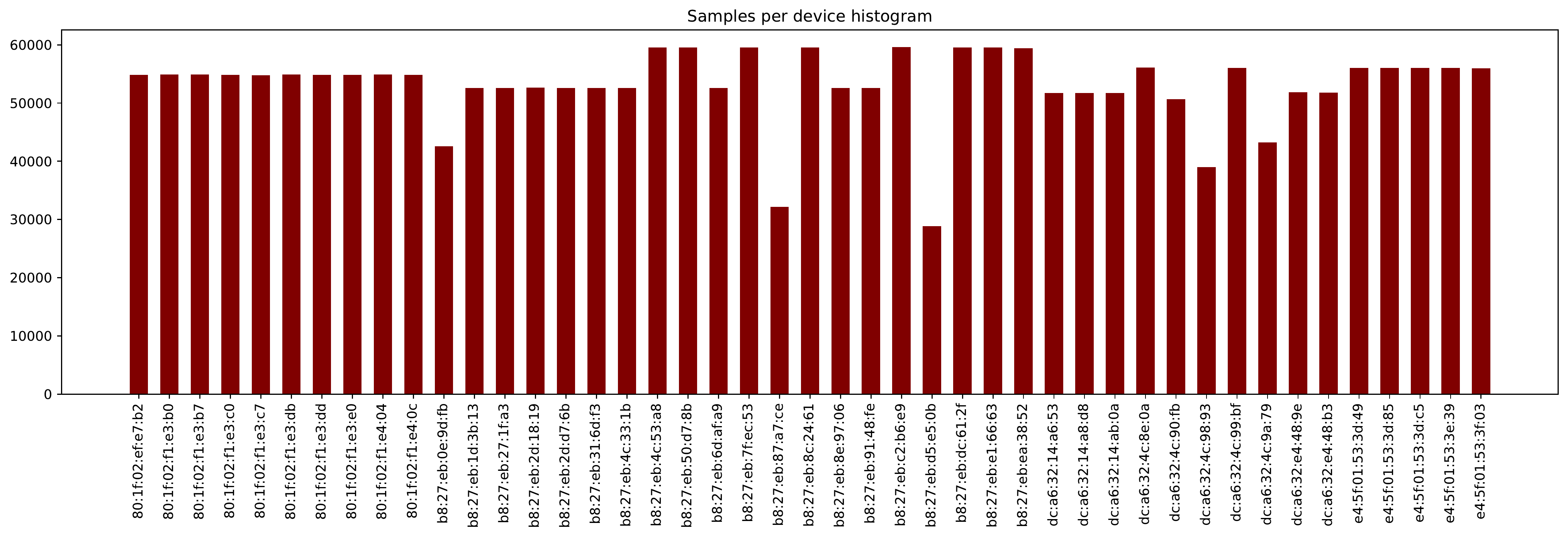}
    \caption{Samples per device contained in the dataset.}
    \label{fig:histogra}
\end{figure*}

A total of 2386126 vectors are available in the dataset, making 4 GB of data. \figurename~\ref{fig:hour_hist} shows the number of samples per hour from each device model during the data collection period. It can also be seen how some devices went offline during the data collection, corresponding to each of the downward jumps shown in the RPi4 and RPi3 graphs. The dataset contains per device model: 505584 samples of RPi 1B+, 784095 samples of RPi4, 547800 samples of RPi3 and 548647 samples of RPiZero. With more than two million vectors, the present dataset is the one with the highest number of samples among those found in the IoT benchmarking literature. Besides, \figurename~\ref{fig:histogra} shows the number of samples per device contained in the dataset. The number varies according to the device model, as more powerful ones generate more data in the same time. Besides, some devices suffered power outages during data collection or they were added lately to the set of available devices. Still, on average, more than 50000 vectors per device are present. It can also be seen that 5 devices have less data due to interruptions during the data collection process.

\section{Data Exploration and Use Cases}
\label{sec:exploration}

This section explores the dataset described in the previous section. For that purpose, a set of ML/DL-based use cases for network and device management are presented using the data available. Note that the purpose of this section is to show the usefulness of the benchmarking application and the data collected with it, not to find the best solution to the problems proposed as illustrative examples.

\subsection{Model/Individual Identification}
\label{sec:identif}

The first use case where the dataset (and LwHBench benchmark to generate new data) can be applied is in device identification based on the performance of its device components. This can be a critical task in environments where the devices can be impersonated by malicious ones with new hardware or software configurations \cite{babun2021cps}. In this sense, this use case can be seen from two different perspectives: \textit{identification of the device model}, a relatively straightforward task from the data measuring the performance of each hardware component, as different hardware will have different performance values; and \textit{individual identification of each device}, a more complex task since to uniquely identify devices with identical software and hardware it is necessary to analyze the differences and imperfections in the chips of each device.

\subsubsection{SBC model Identification}

For this first perspective, a ML/DL-based dimensionality reduction and clustering approach \cite{kriegel2008general} is followed to group together the data from each device model. This approach is selected due to its proven efficacy for automated class inference and identification in unlabeled data applied in many research areas, such as network-based IoT device type inference \cite{sivanathan2019inferring}, intrusion detection \cite{liu2018intrusion} or even biology \cite{granato2018use}.

As a proof of concept, it is decided to apply clustering to all CPU, GPU, memory and storage features, discarding timestamps and temperature. As clustering algorithms, several options are tested, concretely \textit{PCA}, \textit{t-SNE} and \textit{umap}, reducing the number of dimensions to two. So, the resultant data can be easily plotted for result explainability. Regarding clustering, \textit{k-means} and \textit{DBSCAN (Density-Based Spatial Clustering of Applications with Noise)} algorithms are applied over the data once the dimensions have been reduced to two. As the number of device models is previously known, the number of clusters in the algorithm configuration is set to four.

\begin{figure}[htpb!]
    \centering
    \includegraphics[width=0.5\columnwidth]{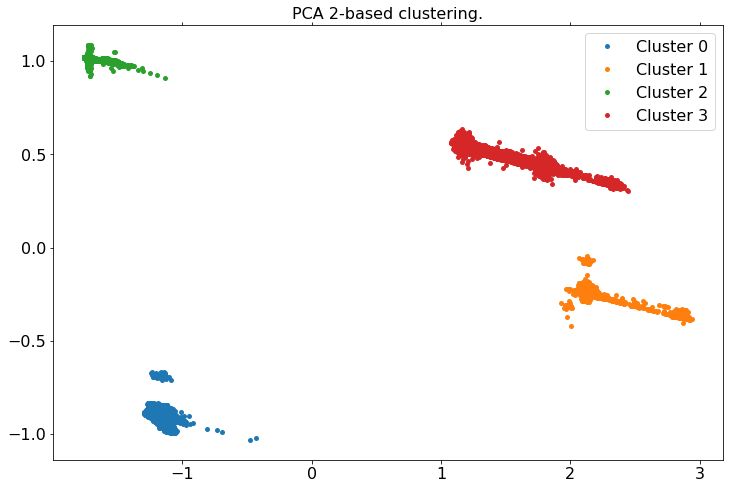}
    \caption{PCA-based dimensionality reduction with k-means clustering.}
    \label{fig:PCA_clustering}
\end{figure}

\figurename~\ref{fig:PCA_clustering} shows the results when \textit{the combination of PCA with k-means clustering} is applied as data processing approach. Different configurations of dimensionality reduction and clustering algorithms gave similar results. It can be seen how four different groups in the data emerge clearly separated, and how the clustering algorithm is able to associate the points successfully. To verify that the clustering is correct, the number of instances assigned to each of them is compared with the number of samples in the dataset belonging to each device model, as also portrayed in \figurename~\ref{fig:hour_hist}. Thus, the results are:
\begin{itemize}
    \item \textit{Cluster 0}: 784095 samples, which coincides with the samples of RPi4.
    
    \item \textit{Cluster 1}: 505584 samples, which coincides with the samples of RPi1.
    
    \item \textit{Cluster 2}: 547800 samples, which coincides with the samples of RPi3.
    
    \item \textit{Cluster 3}: 548647 samples, which coincides with the samples of RPiZero.
    
\end{itemize}

From the above results, it can be concluded that LwHBench benchmark and dataset can be applied to solve the model identification problem. This approach has performed perfectly, since the number of samples in each cluster matches one of the RPi models deployed in the testbed. Moreover, it can be seen how Cluster 2 (green color) has its values much more concentrated than Clusters 0 and 3 (blue and red), which indicates higher stability in the values of this type of device.

\subsubsection{Individual Identification}

In this case, the objective is to uniquely identify each one of the 45 devices used for dataset generation. Here, an ML/DL-based classification approach is followed due to its demonstrated performance in IoT device identification tasks \cite{sivanathan2018classifying, cvitic2021ensemble}.

For this task, all the features available in the dataset regarding components are used, and in this case, it is also included the temperature as a feature, since the correlation between this and the performance of each component can be one of the patterns that the ML/DL algorithm could detect when identifying each device individually. Besides, the storage-related features, 100 measurements for read time and 100 for write, are preprocessed, calculating the average, median, minimum and maximum for each feature group.
As the number of devices to be identified is fixed, the 45 devices used to generate the dataset, ML/DL classification techniques \cite{soofi2017classification} are used to identify each SBC. Therefore, the following techniques are compared: Decision Tree (DT), Random Forest (RF), XGBoost, k-Nearest Neighbors (k-NN), Naive Bayes (NB), Support Vector Machine (SVM) and Multi-Layer Perceptron (MLP). For k-NN, NB, SVM and MLP, normalization is applied using min-max : $x=\frac{x-x_{min}}{x_{max}-x_{min}}$. The dataset is split into 80\% of the data for training and cross-validation, and 20\% for testing. \tablename~\ref{tab:clasif} shows the Precision, Recall and F1-Score \cite{grandini2020metrics} per algorithm. As it can be seen, XGBoost is the algorithm providing the best results, with a 0.97 in Precision, Recall and F1-Score.
\begin{table}[htpb!]
    \centering
    \caption{Device identification results.}
    \label{tab:clasif}
    \begin{tabular}{lcccc}
         \textbf{Algorithm} & \textbf{Hyperparameters} & \textbf{Precision} & \textbf{Recall} & \textbf{F1-Score}  \\
         \hline
         \hline
         DT & $max\_depth=None $& 0.88 & 0.88 & 0.88 \\ 
         \hline
         RF & $n\_estimators=100$ & 0.93 & 0.93 & 0.93 \\
         \hline
         XGBoost & \makecell[c]{$lr=0.1, gamma=0.01$\\$max\_depth=20$}& 0.97 & 0.97 & 0.97 \\
         \hline
         k-NN & $k=7$ & 0.32 & 0.32 & 0.31 \\
         \hline
         NB & - & 0.20 & 0.17 & 0.12 \\
         \hline
         SVM & \makecell[c]{$kernel=linear,$\\$gamma=0.01$} & 0.50 & 0.51 & 0.50 \\
         \hline
         MLP & \makecell[c]{\textit{1 relu hidden layer,}\\ \textit{50 neurons}} & 0.58 & 0.57 & 0.56 \\
         \hline
    \end{tabular}
\end{table}

\begin{figure*}[htpb!]
    \centering
    \includegraphics[width=\textwidth,trim={0 115 0 0} ,clip=true]{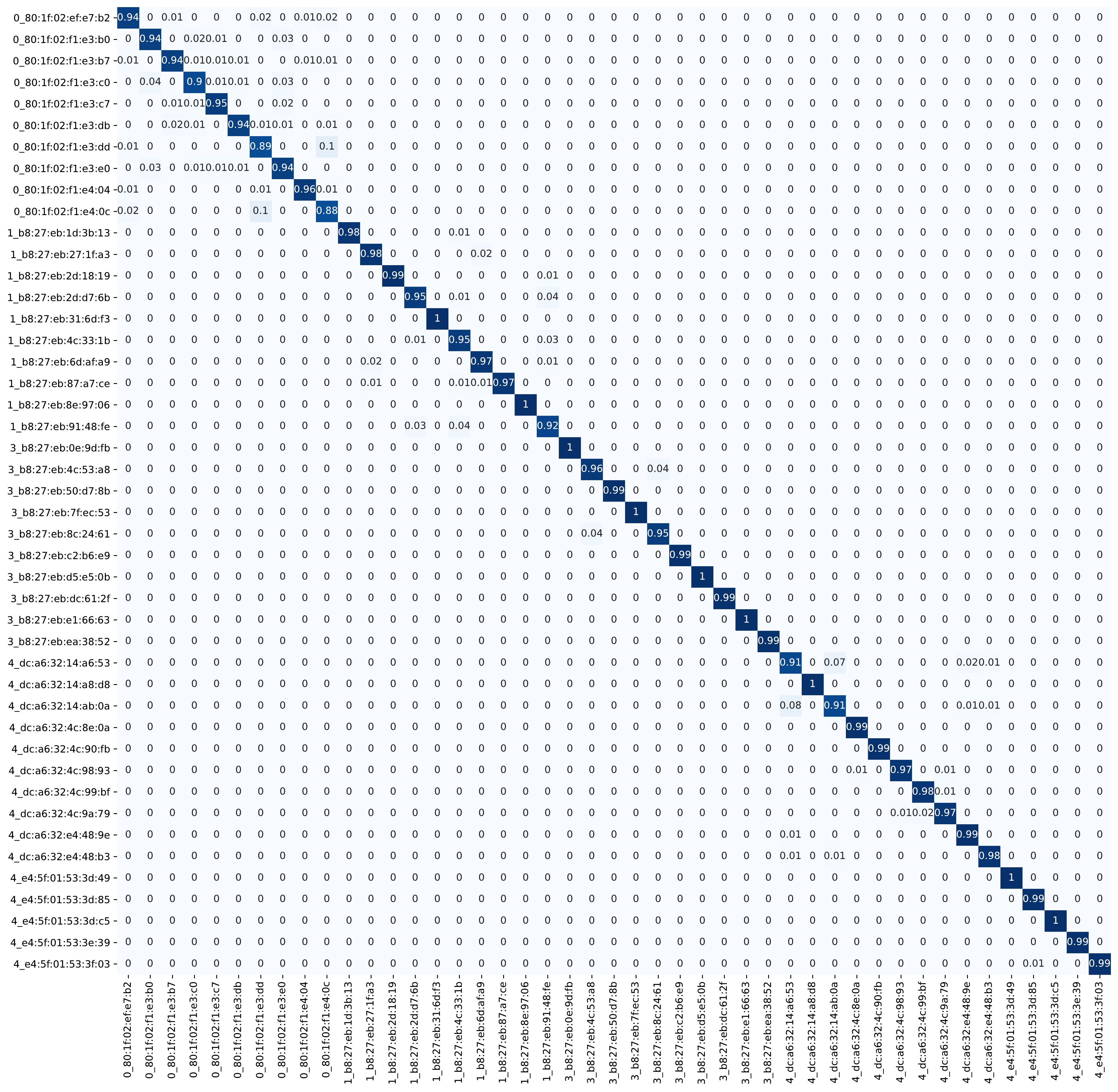}
    \caption{Individual device identification confusion matrix.}
    \label{fig:conf_matrix}
\end{figure*}

Moreover, \figurename~\ref{fig:conf_matrix} shows the confusion matrix for the 45 devices involved in the identification use case. Note that the RPi model has been added at the beginning of each label to have the devices ordered by model in the image. The results in this use case are satisfactory, as the minimum accuracy in all devices is 0.88, having more than 0.93 in most of them. Therefore, it can be concluded that the collected features are suitable for individual device identification.

Furthermore, more complex approaches could be applied to perform individual device identification depending on the scenario requirements. For example, applying time series approaches or advanced DL models such as LSTM (Long Short-Term Memory) or Transformer networks. One example of these solutions can be found in \cite{sanchez2021can}, where similar features to the ones collected about CPU performance were employed to differentiate 25 Raspberry Pi devices. In this case, a sliding window approach is used for vector preprocessing, incorporating new statistical information as features for the ML/DL classifier.

%\subsection{Performance evaluation for a certain task}

\subsection{Performance Analysis}

The second use case where the benchmark and dataset can be applied is in the comparison of the performance of each device hardware component. This comparison can be seen from two different areas: intra-device comparison, where contextual circumstances such as temperature are analyzed to evaluate their impact on the device performance; and inter-device comparison, where components from different devices, but from the same model and in similar context conditions, are compared to find performance variations based on manufacturing variations.

\subsubsection{Intra-device Performance Analysis}

One intra-device use case where the collected dataset can be applied is in the analysis of the impact of temperature variations on the performance of the different device models and their components. This is another research area with a large interest in recent years \cite{gizopoulos2019modern} due to the deployment of SBC in a wide variety of critical scenarios.

For this use case, one of the devices available in the dataset is randomly selected and the impact of temperature on the other hardware-related metrics is analyzed. As the impact of temperature may vary according to the device model, one device per model is selected. Besides, only the first feature regarding storage read and write performance (storage\_read\_1 and storage\_write\_1) are analyzed. In order to test the impact of temperature value on the other features, the correlation between the temperature feature and the other features is studied. The correlation has values between 1 and -1 depending on whether the features increase their value linearly and positively, or inversely and proportionally. 

\begin{figure}[htpb!]
    \centering
    \includegraphics[width=0.5\columnwidth]{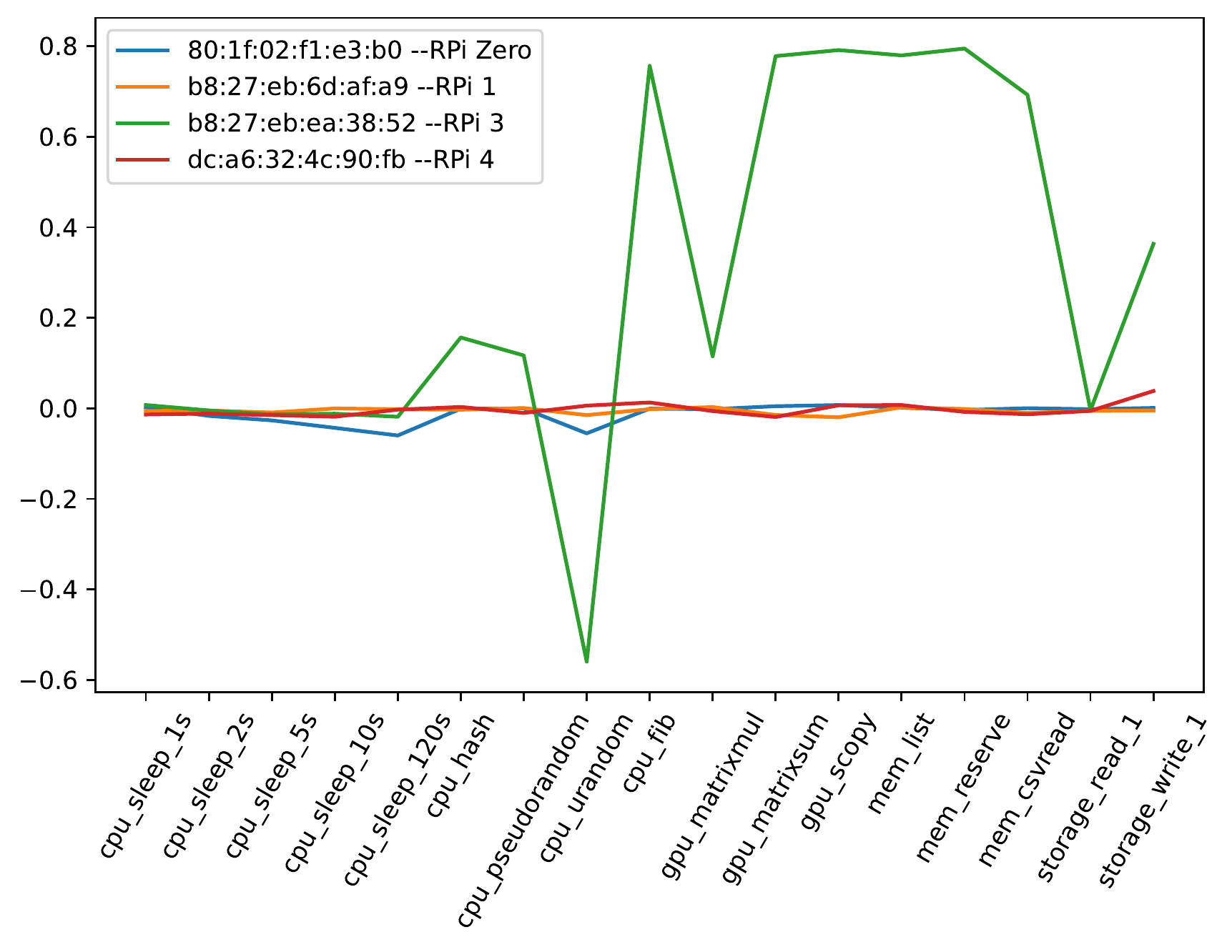}
    \caption{Temperature correlation plot.}
    \label{fig:temp_correlation}
\end{figure}

\figurename~\ref{fig:temp_correlation} shows the correlation values for four different devices, one per available model. It can be appreciated how the RPi3 shows a high sensitivity to temperature in almost all features, only the ones based on sleep function execution seems to have a stable performance. This analysis has been repeated with the rest of the RPi3 devices to ensure that it was not a failure in one of them, with all the devices of this model showing very similar correlation graphs. In contrast, the rest of the models show much lower sensitivity to temperature changes, with values close to zero that only vary around $\pm$0.05 in certain cases and devices.

Thus, from this use case it can be concluded that the RPi3 devices used for the generation of the dataset are the model with the highest sensitivity to temperature changes, as the correlation of this feature with the rest is high in some cases. This fact is interesting for deployments where the physical environment conditions are changing, but the performance is expected to remain stable over time. 

Another interesting use case of intra-device analysis is the analysis of how the performance of a device can drop over time due to component wear and tear. For this use case, data needs to be collected over a long period of time. In the case of the available dataset, a total of 100 days (from 6th December 2021 to 17th March 2022) of data have been collected, so although it is not an extensive period, it could give clues about component and device aging.

\subsubsection{Inter-device Performance Analysis}

The last use case to be explored is the analysis of the performance variations of different hardware components within devices of the same model. This use case is closely related to individual identification, since it is the performance variations between devices of the same model that can be used to characterize each device separately.

To analyze this use case, the distribution densities of different features for devices of the same model are plotted. As a proof of concept, the selected model is the RPi4 and one feature from each component is shown: \textit{cpu\_sleep\_120s} for the CPU, \textit{gpu\_matrixmul} for the GPU, \textit{storage\_read\_1} for the memory/storage, and \textit{storage\_write\_1} for the storage. Although identical experiments could be performed for the other models, these are omitted for document space reasons.

\begin{figure}[ht!]
    \centering
    \begin{subfigure}{0.45\textwidth}
    \includegraphics[width=\textwidth,trim={0 0 0 0} ,clip=true]{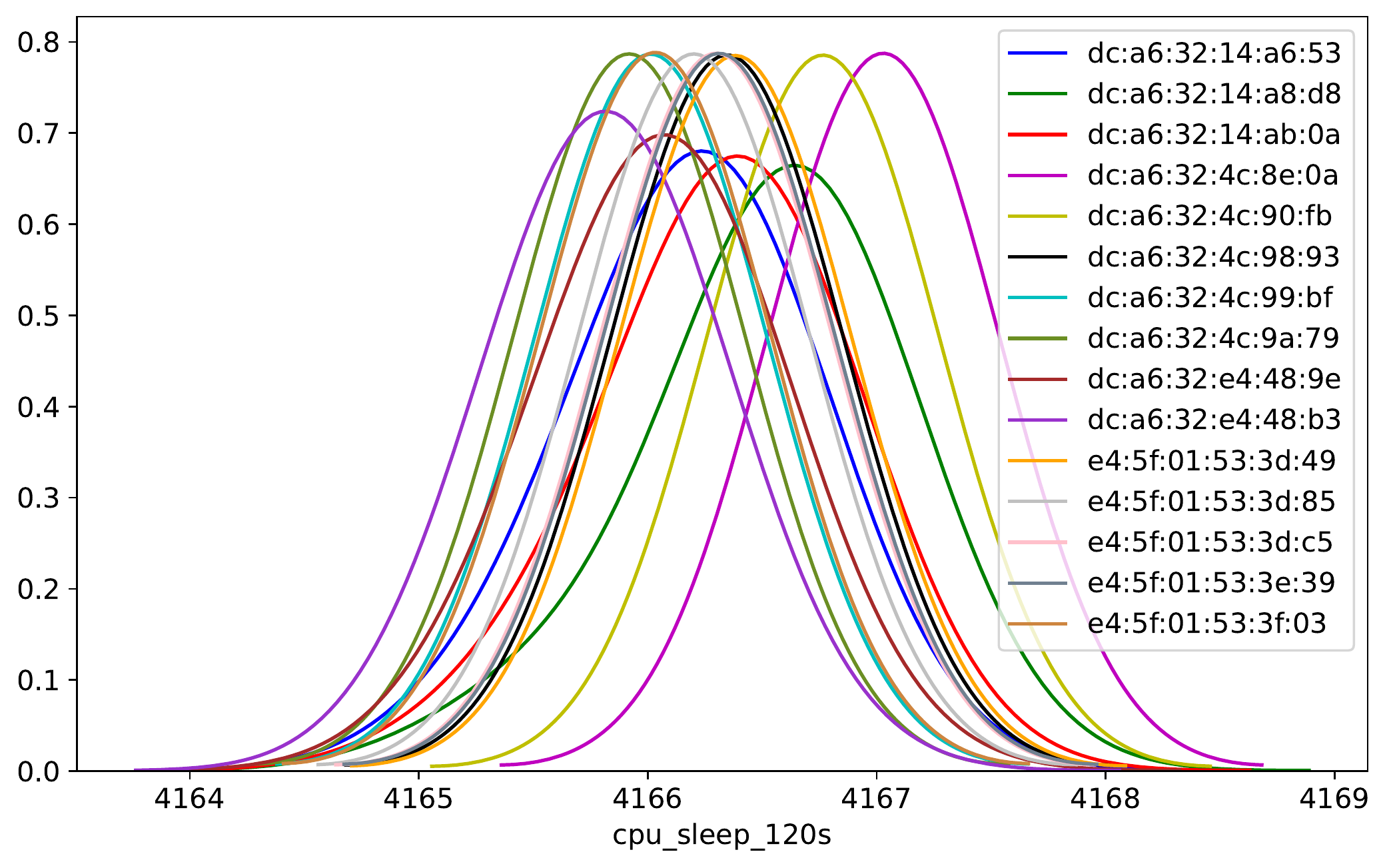}
    	\caption{CPU\_sleep\_120s density plot.}
    	\label{subfig:cpu_sleep_120s}
	\end{subfigure}
	    \begin{subfigure}{0.45\textwidth}
    \includegraphics[width=\textwidth,trim={0 0 0 0} ,clip=true]{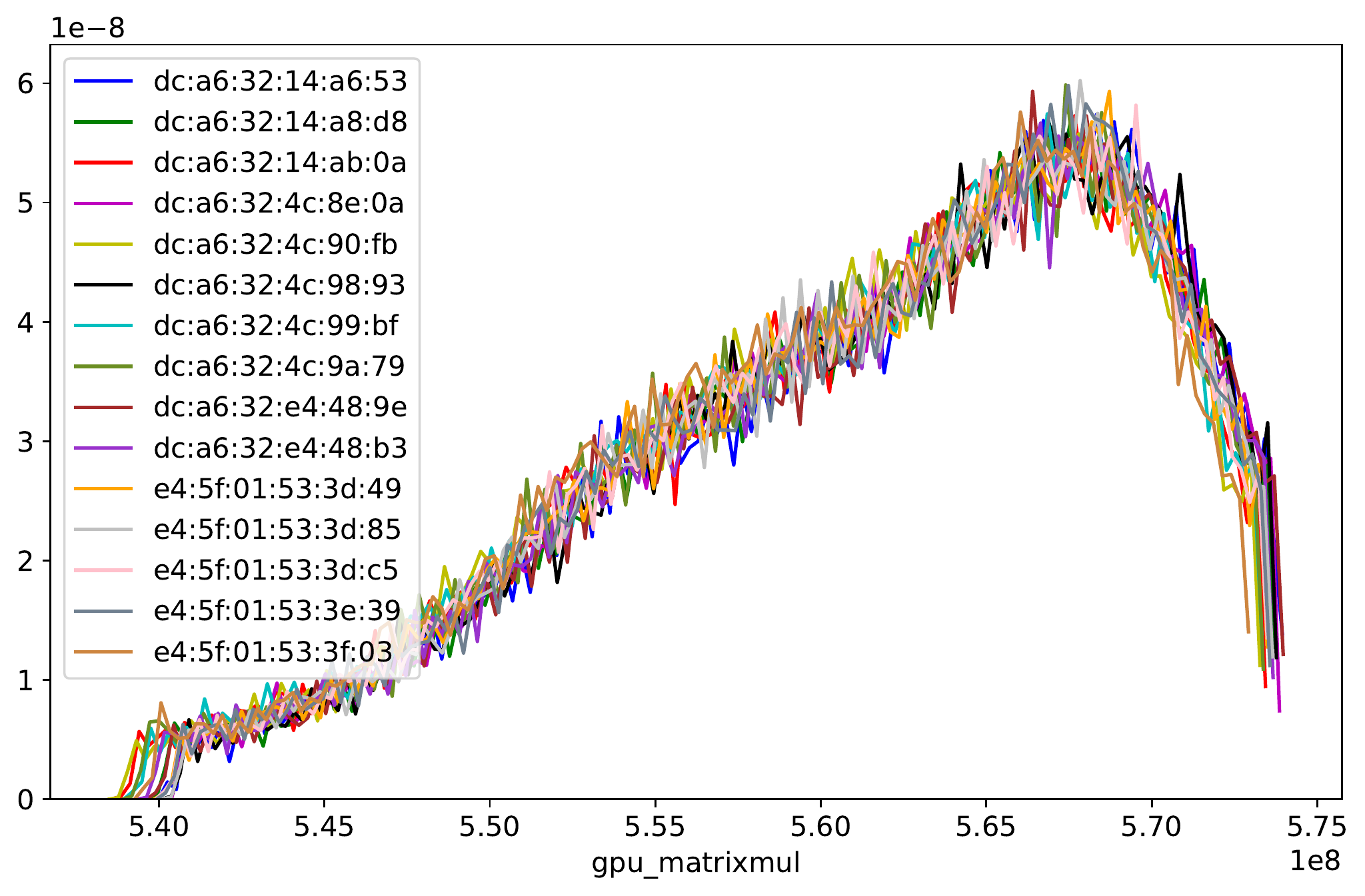}
    	\caption{gpu\_matrixmul density plot.}
    	\label{subfig:gpu_matrixmul}
	\end{subfigure}
	    \begin{subfigure}{0.45\textwidth}
    \includegraphics[width=\textwidth,trim={0 0 0 0} ,clip=true]{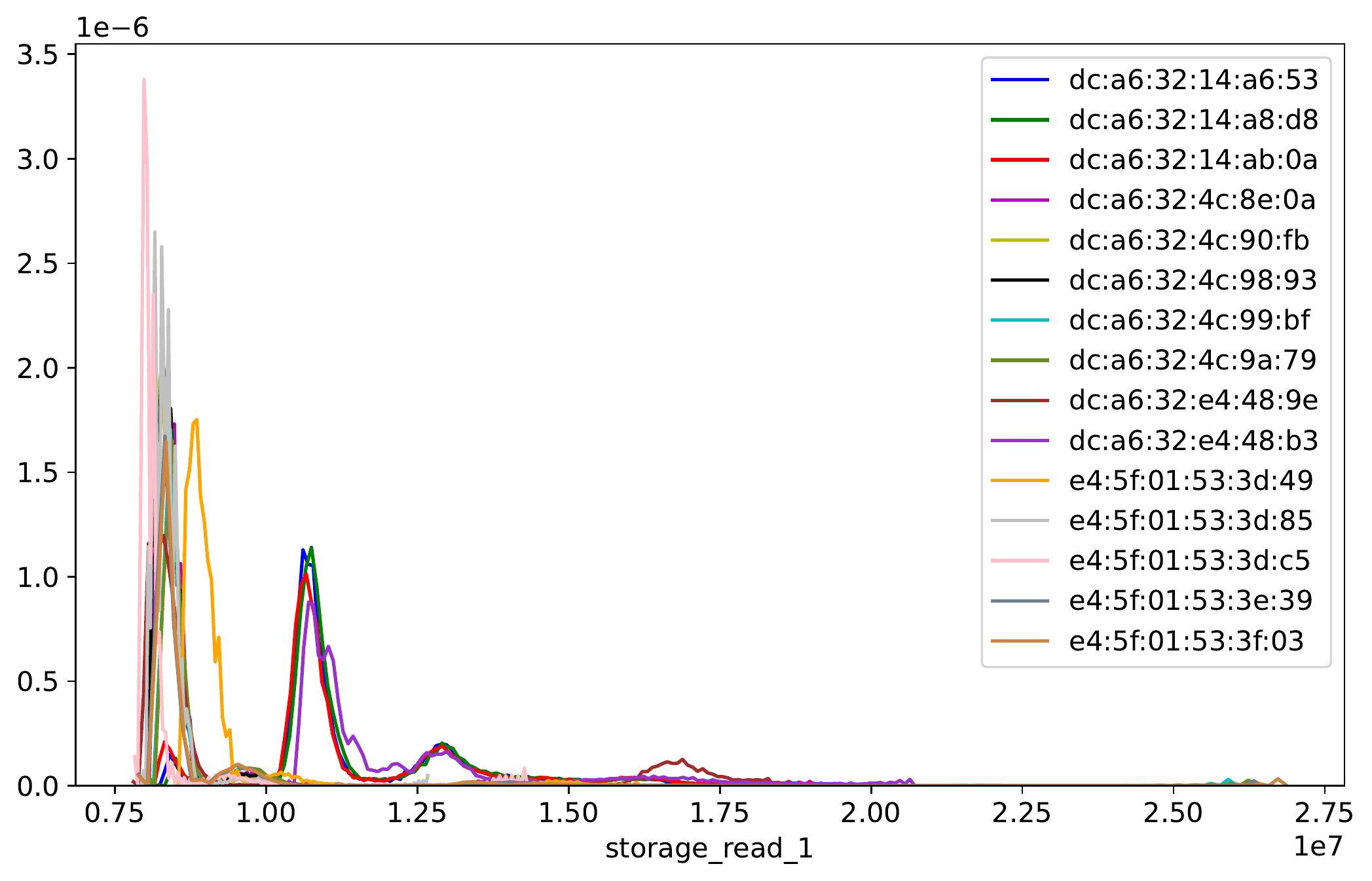}
        \caption{storage\_read\_1 density plot.}
        \label{subfig:storage_read_1}
	\end{subfigure}
	    \begin{subfigure}{0.45\textwidth}
    \includegraphics[width=\textwidth,trim={0 0 0 0} ,clip=true]{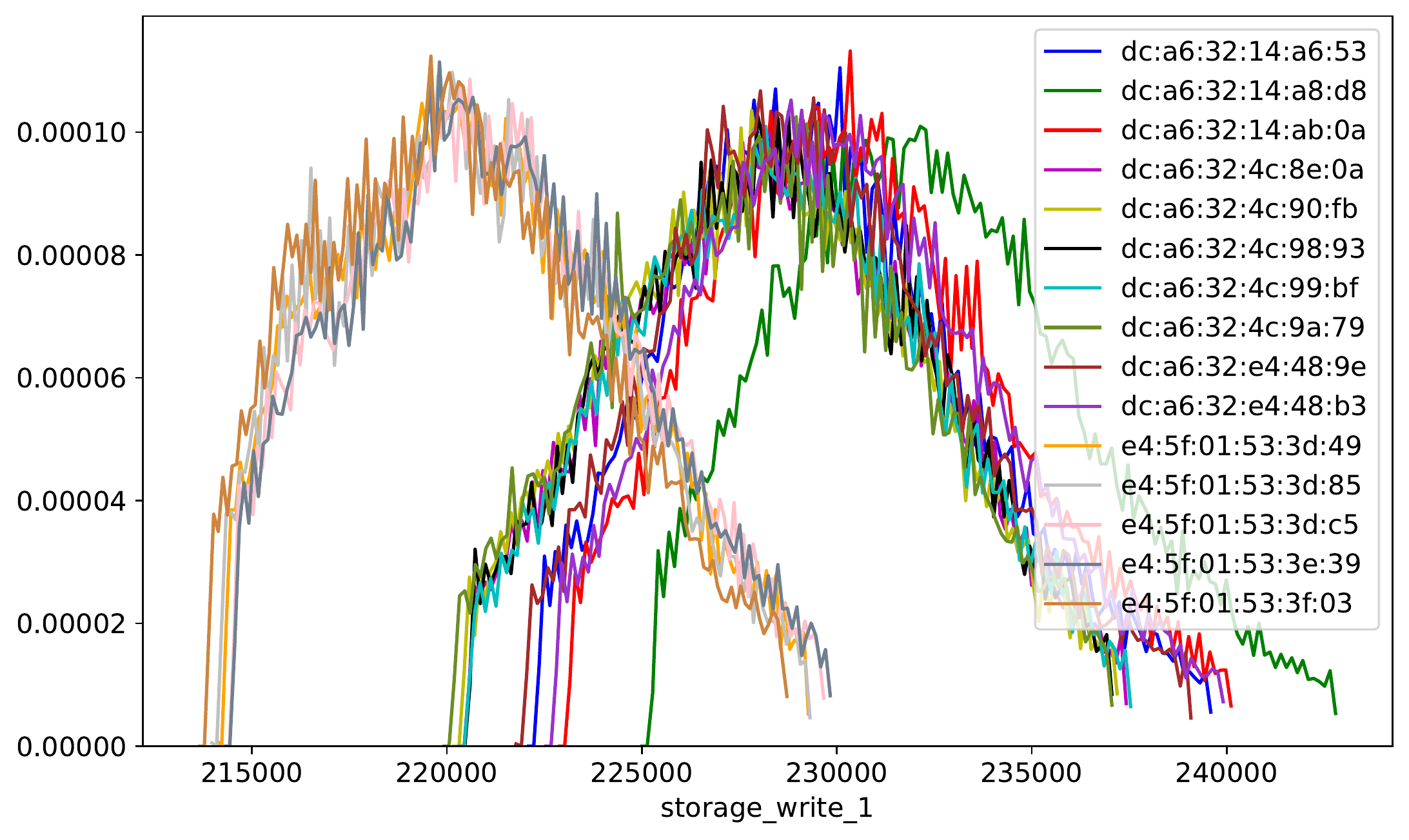}
    	\caption{storage\_write\_1 density plot.}
    	\label{subfig:storage_write_1}
	\end{subfigure}
    \caption{Raspberry Pi 4 feature density plots.}
    \label{fig:inter-device}
\end{figure}

\figurename~\ref{fig:inter-device} shows the density plots for each feature. Some interesting patterns can be appreciated in these plots. First, it can be seen how for the features \textit{gpu\_matrixmul} (\figurename~\ref{subfig:gpu_matrixmul}) and \textit{storage\_read\_1} (\figurename~\ref{subfig:storage_read_1}) the distributions are quite stable between the devices and all of them show similar shapes. However, \textit{cpu\_sleep\_120s} (\figurename~\ref{subfig:cpu_sleep_120s}) exhibits how each device has a Gaussian distribution centered in a different value, demonstrating that some performance variations are present within the device model. %These variations are the ones that are also leveraged in the \textit{Individual Identification} use case to uniquely recognize each device although their hardware specifications are identical. 
A similar situation is found with \textit{storage\_write\_1} (\figurename~\ref{subfig:storage_write_1}) feature, but in this case, the distribution plots tend to have two distinctive center values, and each device sticks to one of them (except for the dark green one with MAC \textit{dc:a6:32:14:a8:d8} that has a higher value, although the distribution shape is similar to the rest).

From this use case, it can be concluded that performance variations are present within each device model depending on the exact device. So, it has been shown that although hardware specifications may be identical, chips contain variations that can be leveraged to perform fingerprinting or identification tasks.

\section{Real-world Deployment in an Agriculture Scenario}
\label{sec:deployment}

This section shows the benchmark adaptation and deployment into a real-world IoT environment based on SBC devices. Here, the aim is to show the use case applicability of the proposed solution as well as its adaptability process to other SBC models. 

In this sense, an IoT sensor network for agriculture has been considered, being the sensors controlled by three PINE64 RockPro64 devices. In this scenario, the administrator wants to keep control of the environment, monitoring the performance of the devices during run time while keeping them identified based on their manufacturing variations. Each device includes a 6-core CPU: 4 x ARM Cortex A53 cores @ 1.4GHz + 2 x ARM Cortex A72 cores @ 1.8 GHz; as GPU it features ARM Mali T860; and 2GB LPDDR4 RAM. As operating system, they use 64 bit armbian (Debian-based Linux for ARM).

As the GPU is different to the ones included in RPi devices, the code needed to be adapted to gather the counters from the ARM Mali T860 GPU. Concretely, the \textit{GPU\_ACTIVE} counter was selected from the ones available \cite{mali_counters}. For cycle counter collection, \textit{ARM HWCPipe} library \cite{HWCPipe} has been employed. For the CPU-based time gathering, \textit{perf} time is gathered in the same way that for RPi devices, using the \textit{perf\_counter\_ns()} function. The functions executed are the same as the ones depicted in \tablename~\ref{tab:features}, adapted in the case of the GPU to the new hardware using the \textit{ARM Compute Library}~\cite{compute_library}. The code is also available in \cite{code}.

\begin{figure}[htpb!]
    \centering
    \includegraphics[width=0.5\columnwidth]{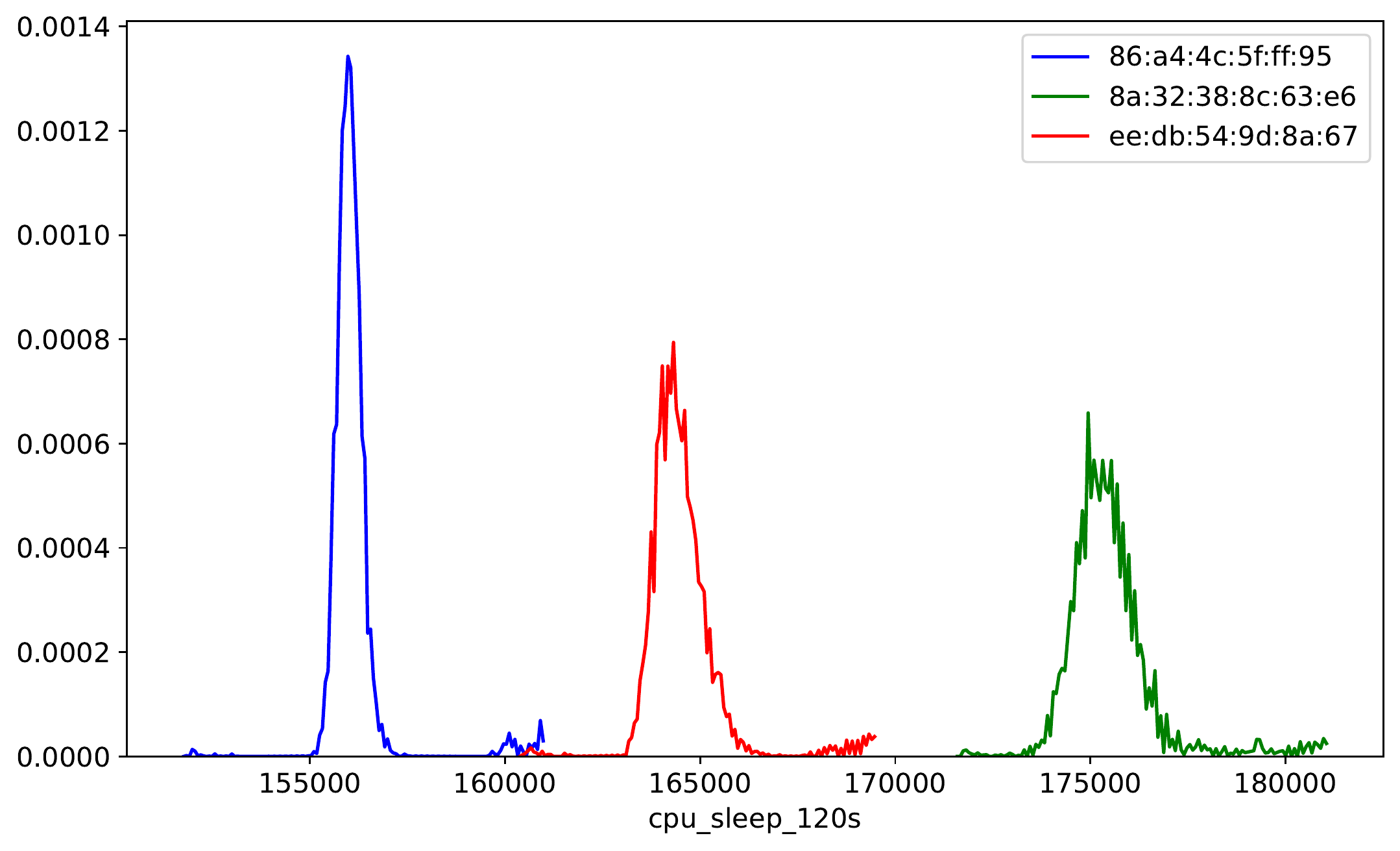}
    \caption{RockPro64 SBC CPU\_sleep\_120s density plot.}
    \label{fig:use_case}
\end{figure}

For experimentation purposes, the code was deployed on the three identical devices during one week. $\approx$35.9 MB of data were collected, with a total of $\approx$12800 vector samples (around 4000 per device). After the data were gathered, the approaches shown in Section \ref{sec:exploration} were implemented to monitor the performance differences between devices and perform individual identification. \figurename~\ref{fig:use_case} shows the distribution plot for the CPU\_sleep\_120s feature. As it can be seen, the three devices distributions are clearly differentiated between each other. Therefore, the classification results, using the same algorithms as the ones depicted in \tablename~\ref{tab:clasif}, are perfect, giving 100\% F1-Score and accuracy.

This real-world deployment demonstrates how the benchmarking application can be adapted and deployed in new SBC models with a relatively low effort, enabling the performance analysis of the devices where the application is deployed. This performance analysis enables different use cases, such as individual identification, to be applied in the scenario.

\section{Discussion}
\label{sec:dicussion}

This section seeks to analyze the main advantages and weaknesses of the proposed benchmarking application and the associated dataset, highlighting the main lessons learned during the work development. From the advantages point of view, it is worth noting the next aspects:

\begin{itemize}
    \item \textbf{An important literature gap has been covered.} As Section \ref{sec:related} shows, there is no low-level benchmarking application for SBC devices, enabling precise hardware analysis, nor any dataset regarding low-level SBC performance. The present work has partially covered these issues with the implementation of the benchmarking application and the release of a large dataset collected for 100 days.
    
    \item \textbf{Demonstrated utility for ML/DL-based use cases.} The collected dataset has been validated in a set of realistic use cases related to AI-based service management, mainly regarding device identification. Thus, it has been shown the utility of the benchmark when it comes to maintain under control an IoT environment where device fingerprinting is critical.
    
    \item \textbf{The benchmark is easy to deploy and extend in other RPi-based environments}. The LwHBench benchmark code is completely open-source \cite{code}, so other researchers and system administrators can execute it in their own scenarios just by installing the required dependencies. This fact also allows for the extension of the benchmark with new metrics according to the requirements of other scenarios or the available hardware.
    
\end{itemize}

Besides, from the drawback perspective, the next points have importance, mainly in future related research:

\begin{itemize}
    \item \textbf{Hardware-based implementation.} LwHBench leverages CPU and GPU cycle counters, which accessed to monitor the performance of other components. Therefore, the implementation for new SBC models may require deep hardware study and understanding. For example, to access to the GPU cycle counter in a new GPU model, it would be needed to read the documentation and try the specific hardware drivers.
    
    \item \textbf{The collected metrics are not directly applicable for traditional benchmarking.} CPU performance is measured in terms of GPU cycles, an unusual metric that is not very representative of the actual hardware performance when executing high-level tasks. Moreover, the differences between devices from the same model can come both from the measured or the measuring device, so it is difficult to measure where the chip imperfections are actually located.
    
    \item \textbf{Component isolation might reduce device performance}. The dataset has been collected while executing LwHBench in an isolated core, when possible (RPi3 and RPi4), and reducing the kernel interruptions from other processes to the maximum. Although the other cores can be used without performance limitations, isolating one core can downgrade the performance of critical tasks running in the SBC at the same time. Therefore, further experimentation with weaker isolation measurements can be interesting to know the impact of other processes on the collected values.
\end{itemize}

This work proposes the only low-level benchmark available in recent literature, and its usefulness has been validated through a series of ML/DL-enabled use cases, which are exciting topics for future IoT-based network and service management solutions. However, as this section details, the proposed solution has room for improvement regarding adaptation to other SBC models.

\section{Conclusions and Future Work}
\label{sec:conclusion}

This work has presented a low-level hardware component benchmarking application for SBC, namely LwHBench. It measures the performance of the CPU, GPU, Memory and Storage of the devices using other self-contained components. This approach ensures that the metrics collected are reliable and non-dependent on the imperfections of the component being measured. The benchmark has been implemented for Raspberry Pi devices, for all the models currently available in the market, from RPiZero to RPi4. In order to ensure optimal measurement stability, every possible action that can help to reduce the noise introduced by other programs running on the device has also been considered, such as isolating the core where the benchmark is running or blocking kernel interrupts.

In addition to the application implementation, an exhaustive dataset has been collected from running LwHBench benchmark on a set of 45 devices over 100 days, which contains more than 2 million vectors and 4 GB of data. Subsequently, to explore the available data and show possible areas of use, a series of use cases have been described and partially solved, reflecting the real needs of an environment whose management integrates modern AI-based solutions. These use cases have been divided into two groups: one on device identification, where it has been shown that using LwHBench it is possible to identify both the model and each device individually; and another on performance analysis, where the possible impact of temperature on hardware performance as well as the variations between devices of the same model have been studied.

As conclusions wrap up, the previous contributions intend to advance state of the art regarding system and device management, as it enables new solutions that, based on low-level hardware benchmarking and ML/DL techniques, improve the control over the devices deployed in modern networking environments, such as 5G-based industries.

As future work, it is planned to adapt the benchmarking application to other SBC models in order to collect data from them and perform more in-depth experiments regarding single-device identification and performance impact of temperature and device aging. Moreover, the benchmarking application will continue being executed in the current devices, generating more data in order to analyze new use cases such as device and component aging. Finally, as further research line, it is planned to integrate the benchmarking application with federated learning techniques, so the data is processed directly in the device according to the use case, without requiring to take the data outside the SBC device.

\section*{Acknowledgment}

This work has been partially supported by \textit{(a)} the Swiss Federal Office for Defense Procurement (armasuisse) with the TREASURE and CyberSpec (CYD-C-2020003)  projects and \textit{(b)} the University of Zürich UZH.

\bibliographystyle{unsrt}
\bibliography{references}

\end{document}